# Study of magnetism in MgO/FeCoB/MgO trilayers using x-ray standing wave techniques

Md. Shahid Jamal[1], Pooja Gupta[2,3], Ilya Sergeev[4], Olaf Leupold[4], Dileep Kumar[1,*]

1. UGC-DAE Consortium for Scientific Research, Indore, India.
2. Raja Ramanna Centre for Advanced Technology, Indore 452013, India.
3. Homi Bhabha National Institute, Training School Complex, Anushakti Nagar, Mumbai 400094, India.
4. DeutschesElektronen-Synchrotron DESY, Notkestraße 85, 22607 Hamburg, Germany.

[*]Email:dkumar@csr.res.in

**Abstract:**

Interfaces in MgO/FeCoB/MgO trilayer have been studied with grazing incident nuclear resonance scattering (GINRS) using the x-ray standing waves (XSW) technique. High depth selectivity of the present method allows one to measure magnetism and structure at the two interfaces of FeCoB, namely, FeCoB-on-MgO and MgO-on-FeCoB, independently, yielding an intriguing result that both interfaces are not symmetric. A high-density layer with an increased magnetic hyperfine field at the FeCoB-on-MgO interface suggests different growth mechanisms at the two interfaces. The azimuthal angle-dependent magneto-optic Kerr effect measurements reveal the presence of unusual uniaxial magnetic anisotropy (UMA) in the trilayer. An in-situ temperature-dependent study discovered that this UMA systematically reduces with temperature. After annealing at 250 °C, the trilayer starts following the standard Stoner–Wohlfarth (SW) model for in-plane UMA. The trilayer becomes isotropic at 450°C with an order-of-magnitude increase in coercivity. The asymmetry at the interfaces is in turn explained by boron diffusion from the FeCoB interface layer into the nearby MgO layer. Stress-induced UMA is observed in the boron-deficient FeCoB layer, superimposed with the bulk FeCoB layer, and found to be responsible for unusual UMA. The temperature-dependent variation in the UMA and coercivity can be understood in terms of variations in the internal stresses and coupling between FeCoB bulk and the interface layer.

## 1. Introduction

FeCoB-based thin films have a significant technological interest as they have the potential for realizing next-generation high-density non-volatile memory and logic chips with high thermal stability and low critical current for current-induced magnetization switching [1]. Over the last few years, the FeCoB-MgO-based layered systems have shown various interfacial phenomena such as interfacial hybridization [1–4], interfacial Dzyaloshinskii-Moriya interaction (iDMI) [5–7], spin-orbit torque (SOT) [8,9], spin Hall effects [10,11], and magnetic anisotropy (MA) [12,13]. The MA and magnetization reversal are the key properties of magnetic thin films for applications and a fundamental understanding of magnetism.

FeCoB alloy has been the most extensively used magnetic electrode with the highest tunnelling magnetoresistance (TMR) among all the magneto tunnel junctions (MTJs) with an MgO barrier [14,15]. In this system, B diffusion is integral to creating a textured FeCo alloy and, consequently, a high TMR value. Many studies claim that B diffuses into the MgO tunnel barrier [16–19], where its presence is suggested to be detrimental to high TMR [20]. Some reports, however, relate the improved TMR to the diffusion of B into MgO [21,22]. This has also been supported by theoretical studies, where the band structure of MgO-B provides a route for coherent tunnelling of electrons, and the diffusion of B is found to be beneficial for the TMR in the MTJs device [23].

Thermal annealing is known to be one of the most crucial steps to achieving high TMR ratios. During the post-annealing process, the TMR increases up to a few hundred percent, from typically 20–40 % in the as-prepared state [14,24–26]. It is suggested that annealing leads to the coherent crystallization of a few monolayers of bcc-FeCo on the MgO, resulting in the large tunneling magnetoresistance due to selective tunneling of spin-polarized electrons of different symmetries. However, recently it has been shown that the spin polarization (SP) of amorphous FeCoB is larger than that of crystalline FeCoB [27]. Thus, there exists some ambiguity about the role of crystallization at the interface in affecting TMR. FeCoB/MgO-based MTJ have also been produced with a large TMR of 240 % using electron beam evaporation. By controlling the evaporation rate, a suitable texture of the MgO layer was achieved. These devices exhibit significantly lower noise than junctions with radio frequency sputtered barriers [28]. Because of the difference in the add-atom energy in sputtering and evaporation, the interface structure in the two cases is expected to be different, influencing the TMR behavior.

In the case of epitaxial Fe thin film, interesting magnetization reversal has been observed when interface-mediated small amounts of uniaxial magnetic anisotropy (UMA) are superimposed on cubic magnetic anisotropy, resulting in single, double, and triple-step hysteresis loops in preferred orientation [29–33]. In several other studies, such unconventional inconsistent hysteresis loops are observed [12,34–36], which do not follow the standard model of magnetic anisotropy. Raju et al. observed multi-jump magnetization reversal in ion beam-sputtered amorphous FeCoB thin films [12] and attributed it to atomic clustering, which provides stable domain states. Recently Sadhana et al. observed a high-density boron-deficient layer at the FeCoB-on-MgO interface [13]. A two-step unusual hysteresis loop in these samples is attributed to the coupling between stress-induced UMA in the FeCoB interface layer and the isotropic bulk FeCoB layer [13]. From the above discussion, it is clear that the interface plays a crucial role in inducing peculiar properties in FeCoB-MgO based systems. Therefore, considerable efforts are also being put together to understand the magnetic and structural properties of the interfaces. But despite extensive investigations [17,18,20,37] it remained ambiguous. Especially the origin of the unusual magnetization reversal and its correlation with interfaces is not yet to be understood. Despite immense interest in this direction, the main difficulty is determining the interface magnetism accurately and correlating the same with magnetic and transport properties unambiguously. Due to the difficulties in getting interface-resolved magnetic information about a thin-layered material, there is an acute lack of systematic studies aiming to understand the observed magnetic properties in terms of the structure of the layers and interfacial regions. In most of the studies in the literature, the conclusion about the interface magnetism, magnetic anisotropy, etc., is extracted based on conventional techniques such as superconducting quantum interface device (SQUID) [38], vibrating sample magnetometer (VSM) [39], magneto-optical Kerr effect (MOKE) [40,41], Mössbauer spectroscopy with $^{57}$Fe probe layer [42] and nuclear magnetic resonance (NMR) [43]. However, these techniques either do not have enough depth resolution so as to resolve the interfaces or may not be probing the real interfaces. Recently x-ray standing wave (XSW) techniques have encouraged experimental efforts to develop interface-resolved and accurate methods, where the x-ray based conventional techniques can be made depth-resolved. It is successfully demonstrated that the depth-resolved information could be obtained using extended x-ray absorption spectroscopy (EXAFS), x-ray fluorescence (XRF), x-ray diffraction (XRD) and x-ray photon spectroscopy under XSW conditions [44–46]. Although these measurements are powerful to provide interface-resolved structural and electronic information in a non-destructive way, the interface-resolved magnetic information of buried interfaces is missing.

In the present study, depth-resolved magnetic and structural properties of MgO/FeCoB/MgO trilayer are studied using synchrotron radiation-based grazing incident nuclear resonance scattering (GINRS) and nuclear resonance reflectivity (NRR) under XSW generated through Pt waveguide structure. GINRS technique made it possible to measure even a fraction of a monolayer of $^{57}$Fe (isotope), whereas by confining x-ray field intensity in waveguide structures, precise magnetic and structural information from the interface is obtained. Observed properties are correlated with the magnetic anisotropy obtained using MOKE by rotating the magnetic field direction in the film plane. The combined analysis is taken to study i) the magnetism at both interfaces, namely FeCoB-on-MgO and MgO-on-FeCoB, ii) the subtle effects of factors like boron diffusion, structural relaxation, and iii) in resolving the existing ambiguities in the literature regarding the role of the interface in magnetic anisotropy.

## 2. Experimental

The trilayer structure MgO(12nm)/FeCoB(10nm)/MgO(6nm) is sandwiched between two Pt layers of 30 nm (buffer layer) and 2.5 nm (capping layer), respectively. These Pt layers form the walls of the planar waveguide to generate XSW in MgO/FeCoB/MgO trilayer. Trilayer behaves as a guiding layer where nodes and antinodes are formed (XSW modes) and provides interface selectivity for x-ray scattering-based measurements. The whole structure was prepared using an ion (Ar$^+$) beam sputtering technique on Si (001) substrate at room temperature (RT) under the base pressure of ~5×10$^{-8}$ torr. A standard target with a composition of $Fe_{43}Co_{40}B_{17}$ was attached with a strip of $^{57}Fe$ (99.999% pure) to enriched FeCoB thin film for isotope-selective measurements. As confirmed using x-ray photoemission spectroscopy measurement, the composition of FeCoB thin film is found to be Fe (46%) Co (39%) B (15%). The chamber was flushed with Ar gas before the deposition to reduce oxygen and water vapor contamination. The deposition rates 0.7, 0.4, and 0.3 Å/s corresponding to the Pt, FeCoB, and MgO layers were kept constant throughout the sample preparation. The sample was annealed at various temperatures such as 150°C, 250°C, 350°C, and 450°C for 30 min at each temperature in a vacuum with a base pressure better than $2\times10^{-9}$ torr.

X-ray reflectivity (XRR) is sensitive to the total thickness of the waveguide structure, including the cavity, the position of the top Pt layer, the interface roughness etc. On the other hand, isotope-selective NRR is particularly sensitive to the $^{57}Fe$-enriched FeCoB layer. Therefore, XRR and NRR measurements were done simultaneously under XSW to characterise the sample structure precisely. In addition, azimuthal angle-dependent hysteresis loops were collected using the MOKE in longitudinal geometry to get magnetic anisotropy in the trilayers [47]. Synchrotron radiation-based GINRS technique is applied to determine the magnitude and direction of magnetic hyperfine fields at the bulk and interface regions of FeCoB. These measurements were carried out at the nuclear resonance beamline P01 at PETRA III, DESY (Deutsches Elektronen-Synchrotron, Hamburg), using energy 14.4 keV ($^{57}Fe$ Mossbauer transition). The synchrotron was operated in the 40-bunch mode with a bunch separation of 192 ns for these measurements. An avalanche photodiode detector (APD) was used for GINRS measurement, having a time resolution of ~ 1 ns. The nuclear (NRR) and electronic (XRR) parts of the signal were separated by making use of the fact that nuclear transitions are delayed in time due to the finite lifetime of the Mössbauer excited state (140 ns in the case of $^{57}Fe$ isotope) [48,49]. Thus, photons detected within a few ns of the incident x-ray pulse constitute the XRR signal due to electronic scattering, while those detected in an interval of 10–160 ns after the incident x-ray pulse are used to get NRR and GINRS patterns.

Depth selectivity of NRR and GINRS are greatly enhanced by generating XSW within MgO/FeCoB/MgO trilayer using Pt waveguide structure. Nodes and antinodes of XSW with within the trilayer are generated through a waveguide structure created by two Pt layers [45]. Resonance modes of such a planar waveguide are excited at the incident angle of x-rays. This satisfies the condition $\theta_i = \theta_m = (m+1)\pi/kW$, where k is the propagation vector of the x-rays, W is the width of the cavity, and m is an integer. Under this condition, the $m^{th}$ transverse electric (TE) mode of the waveguide is excited, which propagates within the Pt waveguide. All GINRS measurements were done at the adequately selected incident angles, where the antinode of XSW coincides at the interface and bulk part of the FeCoB layer. NRR experiments with increasing incident angles were used to select the appropriate angle for interface-resolved GINRS measurements. The depth-profile information about hyperfine interactions in the FeCoB layer and their magnetization direction were studied by fitting the time spectra measured at several grazing angles, which adds reliability to the obtained magnetization depth profiles.

## 3. Theoretical understanding

Isotope selective nuclear resonance technique makes it capable of extracting subtle changes in isotope enriched magnetic layer. The total scattering amplitude ($f_i$) of an atom in the presence of nuclear resonance can be written as a sum of the electronic scattering (responsible for XRR) and nuclear resonance scattering amplitudes. The index of refraction of the layer material in terms of scattering amplitude for grazing incidence geometry is written as [50]-

$$n_{kl} = \sqrt{1 + \frac{\lambda_0^2}{\pi}\sum_i \rho_i (f_i)_{kl}} \qquad 1.$$

At the nuclear transition energies, $f_i$ is consists of scattering by electrons ($f_i^e$) and nucleus ($f_i^n$)

$$n_{kl} \approx 1 + \frac{\lambda_0^2}{2\pi}\sum_i \rho_i [(f_i^e)_{kl} + (f_i^n)_{kl}] \qquad 2.$$

If $\rho_i$ is the volume density of atoms species $i$, $(f_i^e)_{kl}$ and $(f_i^n)_{kl}$ are the electronic and nuclear scattering amplitudes for scattering $\vec{\varepsilon}_l$-polarized radiation into $\vec{\varepsilon}_k$-polarized radiation, respectively. The electronic and nuclear scattering amplitudes (away from the absorption edge, isotropic dipole oscillator and no polarization mixing), are given by-

$$(f^e)_{kl} = -Zr_e + \frac{i\rho_e}{4\pi\lambda_0} \qquad 3.$$

$$(f^n)_{kl} = \delta_{kl} \frac{\lambda_0}{4\pi} \frac{f_{LM}}{1+\alpha} \frac{2j_1+1}{2j_0+1} \frac{A}{x-i} \qquad 4.$$

where $r_e = e^2/mc^2$, Z, and $\rho_e$ are the classical radius of the electron, atomic number, and the electric cross-section, respectively. Where $x = (\Delta E - \hbar\omega)/\Gamma_0$, here A and $\Delta E$ in equation 4 denote inhomogeneous broadening and quadrupole splitting. $\Gamma_0$ is the natural line width, and $f_{LM}$ is the Lamb-Mössbauer fraction. It is clear that the electronic scattering amplitude (electronic contribution to the refractive index- equation-3) does not depend on the energy in the very narrow energy range (~µeV) around the nuclear transition energy.

Thus, the electronic contribution to the refractive index has only a weak dependence on the x-ray energy. Compared to this, the nuclear scattering amplitude for Mossbauer transition (example -$^{57}$Fe nuclei), shown in equation-4, exhibits strong energy dependence around the nuclear resonance (E~14.4 keV). In fact, it dominates the electronic scattering amplitude in this energy region. Therefore, the depth distribution of the isotope (in the present case~ $^{57}$Fe) can be obtained from incident angle-dependent nuclear scattering (NRR). The nuclear and electronic parts of reflectivity were separated by making use of the fact that nuclear transitions are delayed in time due to the finite lifetime of the Mossbauer excited state (140 ns in the case of $^{57}$Fe isotope) [48]. Therefore, photons detected within a few nanoseconds of the incident x-ray pulse (pulse resolution ~0.1 ns) give the XRR (prompt counts) due to electronic scattering, while for nuclear resonance, scattering delays counts are detected for an interval of 10–160 ns. Angular dependence of the integrated resonance counts during the interval 10-160 ns gives NRR pattern, whereas variation in the counts within the delay time (~ 10 to 160 ns) gives time spectra of nuclear resonance scattering (NRS). NRR can be presented in terms of an integral over the resonant counts (delayed count) after prompting by the synchrotron radiation excitation [51]:

$$I^{NRR}(\theta) = \int_t^T I(\theta, t)dt. \qquad 5.$$

Here, T is the interval between synchrotron pulses, and t is the small delay excluding the prompting pulse influence. In above integral $I(\theta, t)$ gives the NRR spectra, and it can be expressed as-

$$I(\theta, t) = \left|\frac{1}{2\pi}\int_{-\infty}^{\infty} R^{\sigma \to \sigma'}(\theta, \omega) \exp(-i\omega t)\, d\omega\right|^2 + \left|\frac{1}{2\pi}\int_{-\infty}^{\infty} R^{\sigma \to \pi'}(\theta, \omega) \exp(-i\omega t)\, d\omega\right|^2 \qquad 6.$$

This intensity expression $I(\theta, t)$ is the Fourier transform of the reflectivity amplitude for σ-polarization of the synchrotron radiation, which is written as-

$$R^{\sigma \to \sigma', \sigma \to \pi'}(\theta, \omega) = \int \chi^{nucl, \sigma \to \sigma', \sigma \to \pi'}(z, \omega) E_\sigma^2(\theta, z, \omega) dz \qquad 7.$$

Where $\chi^{nucl, \sigma \to \sigma', \sigma \to \pi'}$ is the magnetic transverse susceptibility tensor.

It may be noted that compared to the square dependence influence on secondary radiation such as x-ray fluorescence [52], the intensity of the nuclear resonance part from the thin layer is proportional to the fourth power of the standing wave amplitude [51] and hence has more sensitivity compared to the other XSW-based conventional techniques.

## 4. Results

### (i) XRR and NRR measurements

Figure 1(a) gives the XRR as a function of the scattering vector q = 4π sin θ/λ. Along with periodic oscillations (Kiessig fringes) up to higher q values, the XRR pattern exhibits several dips within q= 0.8 nm$^{-1}$ due to increased absorption in the cavity MgO/FeCoB/MgO layer caused by the formation of XSW. This pattern is fitted by taking the thickness of the different layers obtained during deposition using a quartz thickness monitor [45].

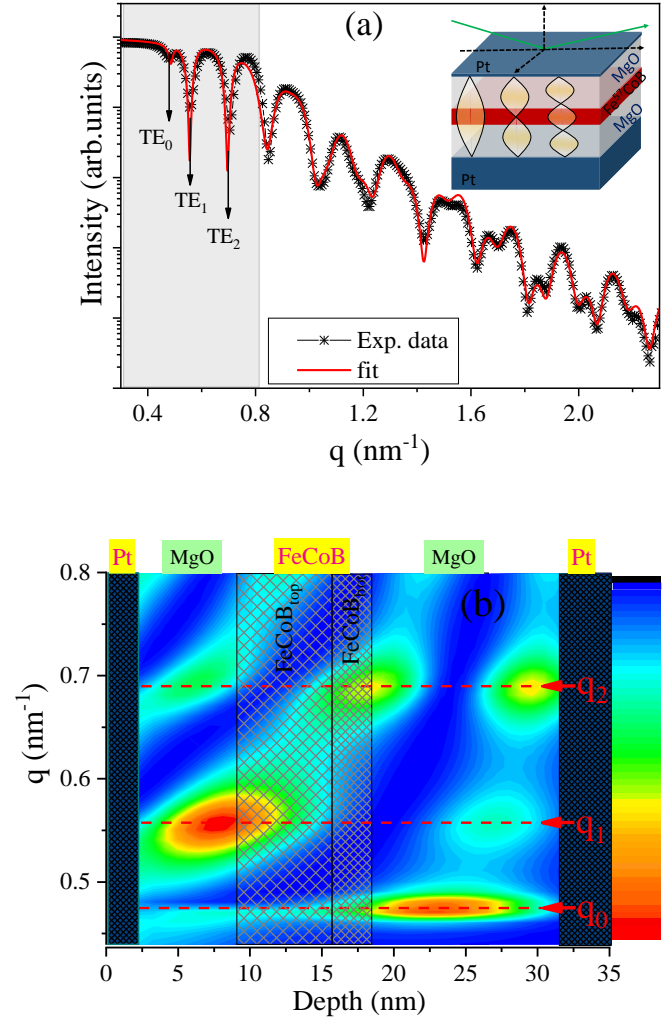

**Figure1**. (a) XRR (symbols) and the corresponding best fit to the experimental data using Parratt's formalism [53]. The inset shows a schematic of the sample structure with different XSW modes distribution along the depth (z) of the WG structure. (b) Calculated x-ray field intensity profile inside Pt waveguide structure. The Positions of Pt and FeCoB layers are marked by a shaded bar. The $q_0$, $q_1$, and $q_2$ are the angles where XSW modes ($TE_0$, $TE_1$, and $TE_2$) are formed.

To get the best fit to the data, it is found necessary to divide the FeCoB layer into two parts of 7.1 nm at the MgO-on-FeCoB side (designated as $FeCoB_{top}$) and 3.0 nm at the FeCoB-on-MgO side (designated as $FeCoB_{bot}$) with higher electron density in $FeCoB_{bot}$ layer. The structure details of the trilayer can be seen in reference [45]. The x-ray field intensity profile in the waveguide structure with increasing q is calculated and plotted in the contour plot shown in figure 1 (b). The position of FeCoB ($FeCoB_{bot}$ and $FeCoB_{top}$) and Pt layers are marked as shaded bars. One may note that as q increases, XSW modes ($TE_0$, $TE_1$, and $TE_3$) are formed at fixed q-values, $q_0 = 0.48$, $q_1 = 0.56$, $q_2 = 0.69$, and $q_3 = 0.84$ nm$^{-1}$.

It may be noted that XRR is sensitive to the total thickness of the waveguide structure, including cavity, Pt layers, interface roughness, etc. But it is not so sensitive to the individual FeCoB layer (morphology, position and width); therefore, fitting the XRR data using FeCoB parameters along with several other parameters may not always lead to genuine information. On the other hand, NRR (nuclear resonance reflectivity) depends on the [57]Fe nuclear transition energy; therefore, it is sensitive to only the [57]FeCoB layer. Thus, the precise and detailed structural information of the total structure is obtained by fitting NRR and XRR curves simultaneously. Figure 2 shows the XRR and the NRR as a function of the scattering vector $q = 4\pi \sin\theta/\lambda$. The q range is covered up to ~ 1 nm$^{-1}$, as all XSW modes in the WG structure are excited within this range. The NRR exhibits well-defined peaks almost at the same q-values, corresponding to each dip in the XRR.

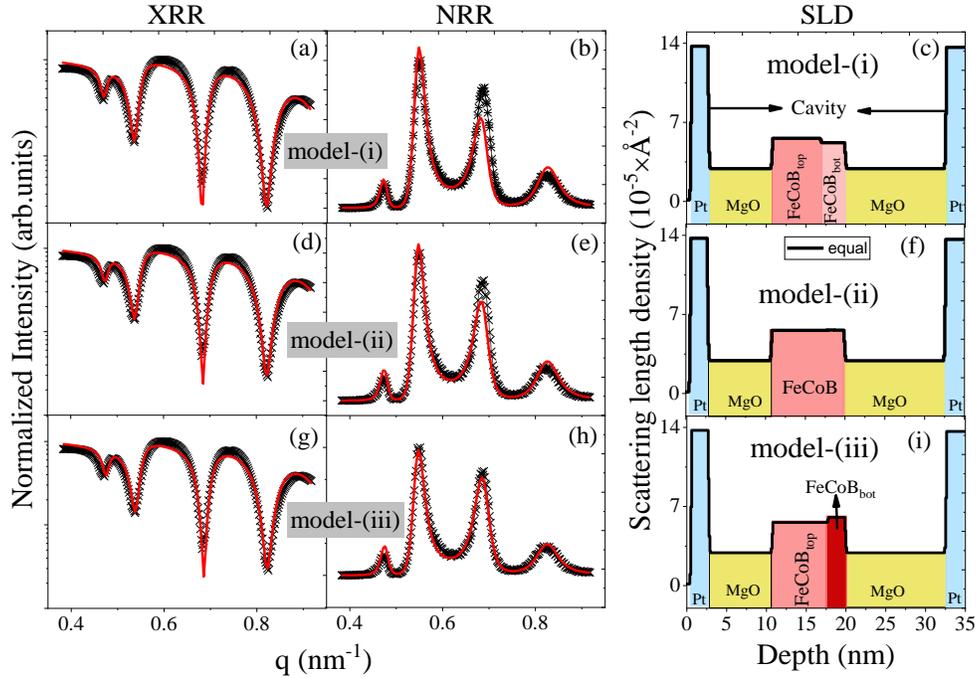

**Figure 2.** (a, d, g) X-ray reflectivity (symbols) and (b, e, h) nuclear resonance reflectivity (symbols) and corresponding simultaneous fits (continuous line) by considering three different models; i) low density of FeCoB at FeCoB-on-MgO interface (c), ii) uniform density of FeCoB throughout the layer (f), and iii) high density of FeCoB layer at FeCoB-on-MgO interface (i).

The origin of these peaks could be understood with the help of the XSW field intensity contour plot in figure 1(b). For incident angle, $q_0 = 0.48$ nm$^{-1}$, the antinode of TE$_0$ mode partly overlaps with the FeCoB-on-MgO interface, giving rise to the 1$^{st}$ peak in the NRR. With a further increase in q, one of the antinodes of TE$_1$ mode overlaps with the MgO-on-FeCoB interface and gives rise to the 2$^{nd}$ peak. The 3$^{rd}$ and 4$^{th}$ NRR peaks can be understood similarly. The shape of the NRR 3$^{rd}$ peak is sensitive to the variation of $^{57}$Fe concentration across the bottom side of the FeCoB layer, while the 2$^{nd}$ peak depends upon the $^{57}$Fe concentration profile across the top side of the FeCoB layer. It demonstrates that even a small variation in the position and density of the FeCoB layers would result in a significant variation in the intensity ratio of the peaks because, at this depth, the distributions of the XSW antinode have steep gradients. Thus, fitting the NRR peaks provides a sensitive way to study the FeCoB layer.

The NRR data in the right panel of figure 2 (Fig. 2h) is fitted simultaneously with XRR data (Fig. 2g) by taking the thickness obtained by the thickness monitor. To best fit the data, it is found necessary to take a 3 nm thick high-density FeCoB layer towards the bottom interface (FeCoB$_{bot}$). For the comparison, the simulated XRR and NRR curves are also shown in figure 2(a-f) by considering i) a low-density FeCoB layer at the interface and ii) the uniform density of FeCoB throughout the layer. There is a clear disagreement between these simulated curves and experimental data with these two models. The quantitative structural information based on the combined simultaneous fitting confirms two layers within the FeCoB layer; i) d$_{top}$ = 7.1 nm, $\rho_{top}$ = 5.28×10$^{-5}$Å$^{-2}$; ii) d$_{bot}$ = 3.0 nm, $\rho_{bot}$ = 5.89×10$^{-5}$Å$^{-2}$ FeCoB$_{bot}$ (near FeCoB-MgO interface). $\rho_{bot}$ is about 12% more as compared to $\rho_{top}$, and the final sample structure is Si/Pt (33.2 nm)/MgO (11.3 nm)/FeCoB$_{bot}$(3.0 nm)/FeCoB$_{top}$(7.1 nm)/MgO(6.5 nm)/Pt(2.5 nm).

**(ii) GINRS measurements under XSW condition; interface-resolved magnetism**

Since the XSW antinodes at q$_1$ = 0.56 nm$^{-1}$ and q$_2$ = 0.69 nm$^{-1}$ overlap with the bottom and top interface side of the FeCoB layer, GINRS measurements will have interface-weighted magnetic information of the FeCoB layer. Schematic in Fig. 3a gives the geometry and GI-NRS curves obtained at q$_1$ = 0.56 nm$^{-1}$ and q$_2$ = 0.69 nm$^{-1}$are shown in Fig. (3b). In figure (3a), k$_i$ and k$_f$ denote the incident and reflected wave vectors, α$_i$ and α$_f$ represent the incidence and reflected angle of the synchrotron beam. To get the hyperfine field (Bhf) and distribution of the hyperfine field around the mean value (ΔBhf), both curves are fitted simultaneously using REFTIM software [54] by taking trilayer structural as obtained from XRR and NRR measurements. The best fit to the data is obtained by considering three different hyperfine fields (Bhfs) 29.5, 33.2, and 30.0 T in FeCoB$_{top}$ and FeCoB$_{bot}$ layers. The density concentration of all the Bhfs in FeCoB is shown in Fig. 3(c). The broad ΔBhf 5.5, 2.3, and 2.6 T corresponding to the Bhf 29.5, 30.0, and 33.2 T, respectively, confirm the amorphous nature of FeCoB [47]. The cause of different Bhfs contributions in the top and bottom part of the FeCoB layer is due to the different magnetism caused by compositional differences [55,56].

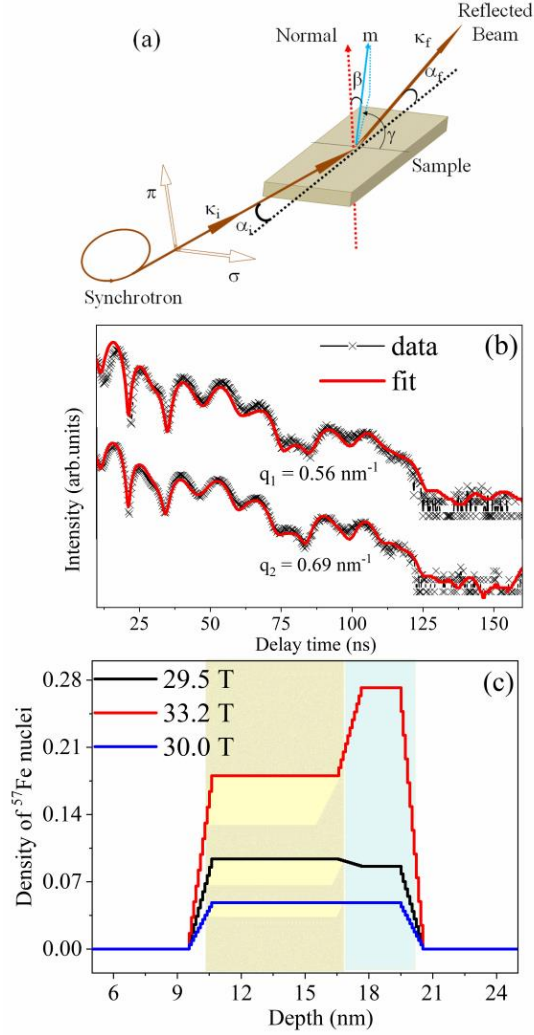

**Fig. 3:** (a) Schematic of the GINRS measurements. ($\beta$, $\gamma$) defines the relative orientation of the magnetization **m** of the sample and ($\sigma$, $\pi$) are the linear polarization basis vectors. (b) GINRS time spectra measured at grazing angles at $q_1$ = 0.56 & $q_2$ = 0.69 nm$^{-1}$ before the critical angle of Pt. (c) The depth distribution of the $^{57}$Fe nuclei with the contribution of different hyperfine fields.

**Table 1**. The Bhfs, $\Delta$Bhf, and their percentage are presented for the as-deposited along with annealed stage at 300 °C and 450 °C samples.

| FeCoB layer between MgO layer | Sample states | d (nm± 0.1) | Bhf (T± 0.5) | ΔBhf (T± 0.1) | Percentage |
|---|---|---|---|---|---|
| FeCoB$_{Top}$ |  | 7 | 29.5 | 5.5 | 14% |
|  |  |  | 30.0 | 2.3 | 7% |
|  |  |  | 33.2 | 2.6 | 27% |
| FeCoB$_{bot}$ | RT | 3 | 29.5 | 5.5 | 13% |
|  |  |  | 30.0 | 2.3 | 7% |

| | | | 33.2 | 2.6 | 31% |
| --- | --- | --- | --- | --- | --- |
| **FeCoB (single layer)** | 300 °C | 9.6 | 30.5 | 3.1 | 12% |
| | | | 31.2 | 2.4 | 16% |
| | | | 35.9 | 1.6 | 72% |
| **FeCoB (single layer)** | 450 °C | 9.1 | 28.7 | 3.1 | 8% |
| | | | 30.3 | 2.7 | 5% |
| | | | 36.3 | 1.9 | 87% |

### (iii). Angular-dependent magnetism using MOKE; magnetic anisotropy

Hysteresis loops were collected using MOKE after applying magnetic fields in different in-plane azimuthal directions ($\varphi$). Fig. 4(a-d) shows some representative loops obtained in $\varphi = 0°, 30°, 60°$, and $90°$ directions. There is a strong variation in the shape of the loops with angle $\varphi$. The hysteresis loop is almost square for the applied magnetic field along $\varphi = 0°$ direction, suggesting that the magnetization occurs through domain wall motion [57]. With the increasing azimuthal angle, the rounding off of the hysteresis curve indicates the increasing contribution of the rotation of domain magnetization. The normalized remanence ($Mr/Ms$) and coercivity ($Hc$) variation, as observed in angular dependence, are shown in Fig. 4(e & f). The dumble-shaped variation of $Mr/Ms$ indicates the presence of uniaxial magnetic anisotropy, whereas the easy and hard direction of magnetization can be understood with the standard Stoner-Wohlfarth (SW) model [58] for in-plane magnetic anisotropy.

According to the SW model, "the minimizing of the free energy of the system in the presence of an applied field H making an angle $\varphi$ from the easy axis can be written as [57]-

$$2m(1-m^2)^{1/2} \cos 2\varphi + \sin 2\varphi \, (1-2m^2) \pm 2h(1-m^2)^{\frac{1}{2}} = 0 \qquad 8.$$

Where $h = H/H_a$ is the reduced field, $H_a$ being the anisotropy field, and $m = M/M_s$ is the reduced-magnetization.

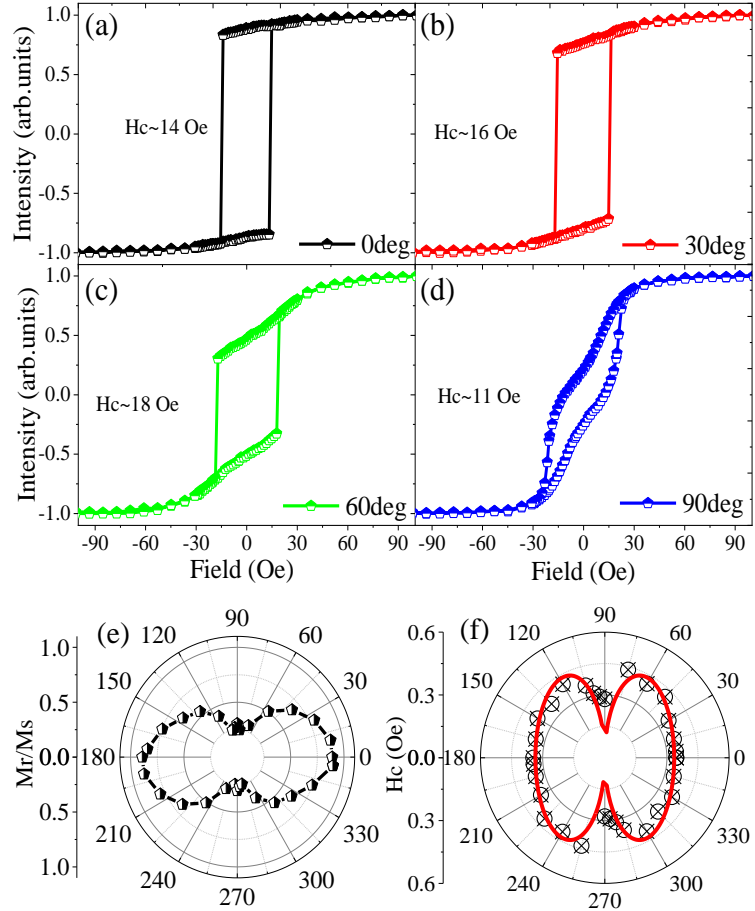

**Figure 4:** (a-d) Representative MOKE loops taken for φ = 0°, 30°, 60°, and 90°. (e) and (f) are the polar plots of *Mr/Ms* and *Hc* as a function of φ, respectively.

The solutions of this equation for different values of $\varphi$ (0°, 60°, and 90°) are shown in the simulated plot in Fig. 5. The m vs h curve for $\varphi = 0°$ is a perfect square (*Mr/Ms*~1) while for $\varphi = 90^0$ the curve is almost linear (Mr/Ms<<1). It is important to note that for intermediate values of $\varphi$, the applied field needed to saturate the magnetization is significantly higher than that needed to saturate the magnetization for $\varphi = 0^0$. Therefore, it is clear that in Figure 5.-

(i)    Square loop (*Mr/Ms*~1) along $\varphi = 0^0$ is the easy axis of magnetization

(ii)    For *Mr/Ms*<<1 or normal to the easy axis ($\varphi = 90^0$) belongs to the hard magnetization axis

The SW model considers coherent rotation of magnetization, and therefore, as shown in the simulated Fig. 5, it predicts a monotonous decrease of *Hc* with increasing φ from 0° to 90°. However, in the present case, the coercivity first increases as $\varphi$ it rotates away from the easy axis ($\varphi = 0°$) but decreases when approaching the hard axis ($\varphi = 90°$), as shown in Fig. 4(f). Therefore, it can't be interpreted using the SW model.

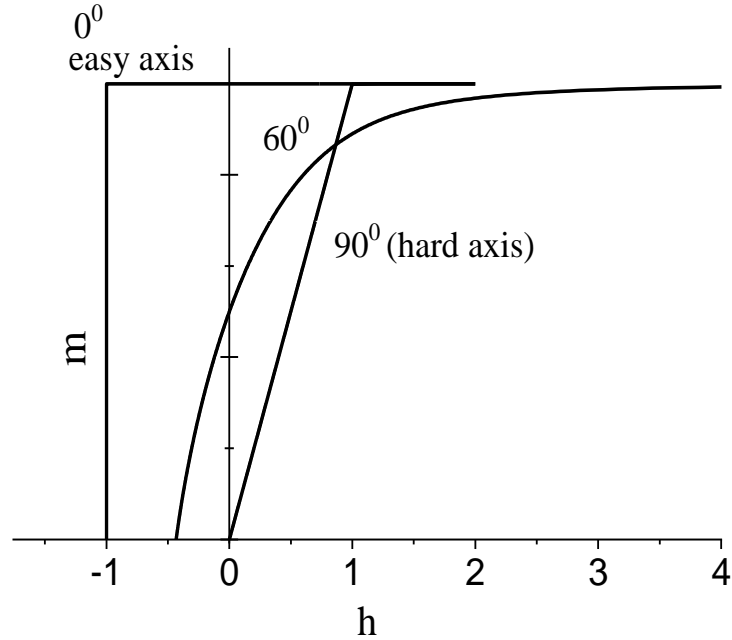

**Figure 5:** Simulated m vs h for angles $\varphi=0°$, 60° and 90° ($\varphi$ is the angle when the applied field is along the direction of the easy axis of magnetization) using Stoner Wohlfarth model of magnetic anisotropy in thin films.

*Hc* variation is understood by including both coherent rotation and domain wall nucleation (two-phase model) and employed to understand the magnetization reversal in the present case. The angular variation in coercivity, in this case, is described as [59].

$$H_C(\varphi) = H_C(0°)\frac{(N_x + N_N)\cos\varphi}{N_z \sin^2\varphi + (N_x + N_N)\cos^2\varphi} \qquad 9.$$

Where $N_z$ and $N_x$ are the demagnetizing factors along $\varphi=0°$ and $\varphi=90°$, respectively. The $N_N=H_a/M_s$ is an effective demagnetizing factor; $H_a$ and $M_s$ are the anisotropy field and saturation magnetization. If $(N_N+N_x)/N_z$ is close to zero, the magnetization reversal mechanism is dominated by the coherent rotation. For an infinite value of this ratio, the magnetization reversal mechanism is mediated by the domain wall nucleation. As given in Fig. 4(f), the angular dependence of coercivity is fitted by equation 9. Two kinds of phases at a low field regime have been employed to account for the angular-dependent magnetization in the FeCoB layer [60]. The angular variation *Hc* is well-fitted with a two-phase model.

### (iv). Effect of thermal annealing

The sample is annealed at various temperatures to investigate the angular-dependent magnetism. Figure6 represent the hysteresis curves obtained along $\varphi=0°$, 80°directionsafter annealing at different temperature of 150 °C, 250 °C, 350 °C, and 450 °C for 30 min. The *Hc* and *Mr/Ms* are plotted as a function of $\varphi$ in the same figure. *Mr/Ms* Vs $\varphi$ for 150°C shows uniaxial anisotropy, whereas *Hc* variation is almost independent of the angle. It is important to note that compared to the as-prepared sample, where *Hc* near the hard axis was higher, at this temperature, it becomes almost the same in both directions. At 250°C temperature, anisotropy in *Hc* and *Mr/Ms* decreased but started to follow the standard uniaxial magnetic anisotropic trend as per the SW model. At this temperature, *Hc* and *Mr/Ms* Vs $\varphi$ plots exhibit decreasing trends with increasing $\varphi$ from 0° to 90°. Anisotropy almost disappeared at a temperature of 350°C. After annealing at 450°C, the *Hc* values dramatically increased, and anisotropy vanished.

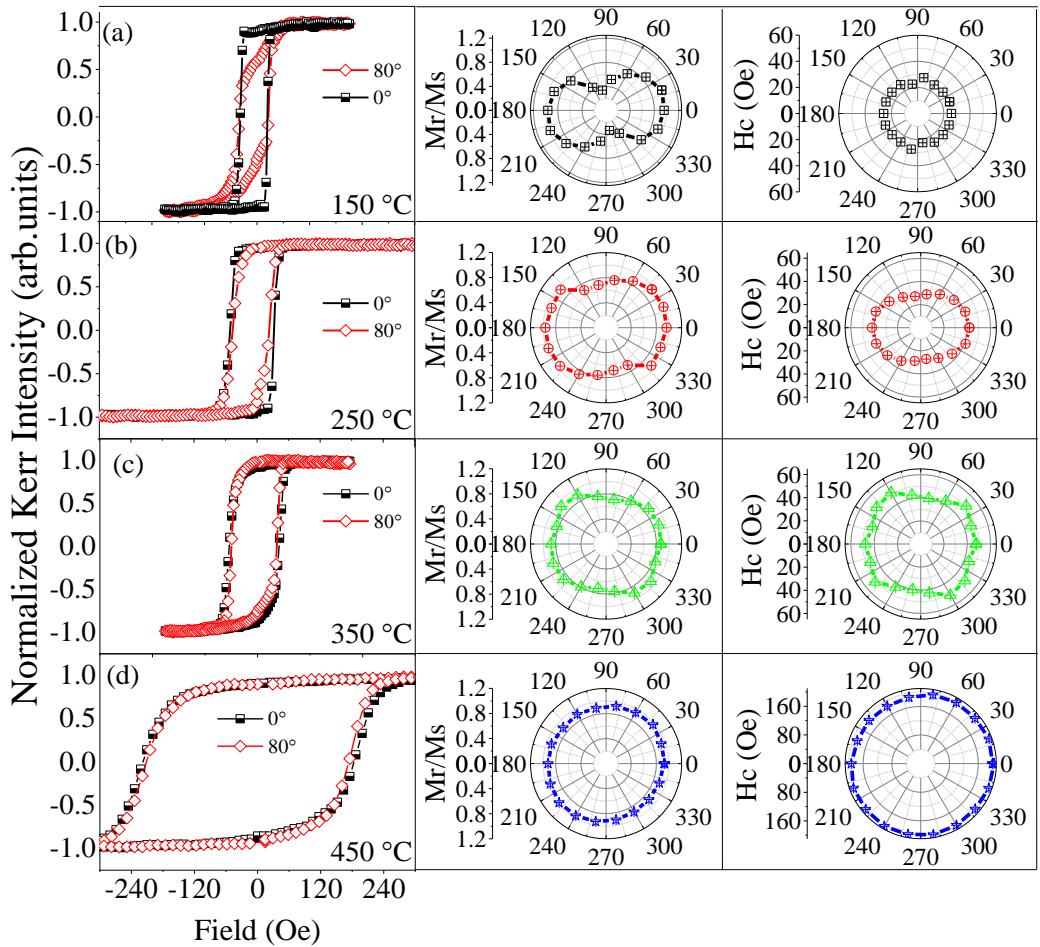

**Figure 6:** Representative MOKE loops with azimuthal angle $\varphi = 0°$ and 80° for annealed samples at (a) 150 °C, (b) 250 °C, (c) 350 °C, (d) and 450 °C.(e) and (f) are polar plots of coercivity (*Hc*) and the squareness ratio (*Mr/Ms*) as a function of $\varphi$, respectively.

Fig. 7(a, b) gives GINRS time spectra measured at grazing angles $q_1 = 0.56$ & $q_2 = 0.69$ nm$^{-1}$ after annealing samples at 300 °C and 450 °C, respectively. The amplitude of the beating pattern is enhanced and more pronounced after annealing with respect to the as-deposited sample. As compared to the as-prepared sample, the experimental data are fitted by considering a single FeCoB layer with three different Bhfs 30.5, 31.2, & 35.9 T and 30.3, 36.3, & 28.7 T corresponding to temperatures 300 °C and 450 °C, respectively. These Bhfs contributions have also been shown in table 1 for comparison. After annealing at 300 °C and 450 °C, Bhfs contribution corresponding to the 35.9 T and 36.3 T is found to be 72% and 87%, respectively. Higher Bhfs (35.9 T and 36.3 T) after annealing in the magnetic layer can be attributed to the crystallization of FeCoB. It is in accordance with some earlier studies, where the crystallization in FeCoB starts due to the diffusion of boron (B) from FeCoB [61,62].

There are various reports on FeCoB crystallization and migration of B into MgO. Using x-ray photoelectron spectroscopy, J. C. Read et al. elucidate that when the CoFeB/MgO bilayer was heated above 350 °C, B moved into the MgO matrix and formed a composite MgB$_x$O$_y$ layer [16]. Another similar report by A. A. Greer et al. studied the distribution of B in a Ta/Co$_{0.2}$Fe$_{0.6}$B$_{0.2}$/MgO sample annealed at 300 °C with standing-wave hard x-ray photoemission spectroscopy (SW-HXPS). They suggested that the diffusion of 19.5% of the B uniformly into the MgO layer and of 23.5% into a thin TaB interface layer [3].

## 5. Discussions

In several studies in the literature, unusual angular-dependent magnetism is always seen in the FeCoB and MgO-based systems and is always attributed to the uniaxial magnetic anisotropy, which does not follow the standard Stoner-Wohlfarth model [34,36,63–65] of magnetic anisotropy. For example, Kipgen et al. observed MA in ion beam sputtered FeCoB on Si(001) substrate and argued that this MA is strain-induced where the crystallographic orientation of the underlying Si(001) substrate played an important role in the anisotropy of the film [35]. In another report, A. T. Hindmarch et al. demonstrated that in-plane MA in amorphous FeCoB thin films grown on epitaxial GaAs(001) substrate [65]. They have attributed the origin of MA is bond orientational anisotropy due to interface interaction. V. Thiruvengadam et al. grown FeCoB on SiO2 and MoS2 substrate [66], MA in the case of only FeCoB/MoS2 is observed and attributed to the Hybridization between MoS2 and Fe or Co. Our observation related to the asymmetry at FeCoB/MgO and MgO/FeCoB interface is in accordance with the theoretical study, where PMA found in FeCoB/MgO/FeCoB structures, driven by FeCoB/MgO interface [1,67].

In the present case, combined data analysis suggests the existence of a 3 nm thick B deficient FeCoB layer at the interface. Hence both parts of the FeCoB layers are expected to be magnetically different because of B diffusion from the interface layer, which suggests stress in the interface layer [13]. B diffusion into MgO at the bottom interface could be due to the porousness of the MgO layer, which might have been generated during the deposition of ion beam sputtering [13]. Due to the small size of B compared to the Fe and Co and the high electron affinity of Mg and O, B has a great chance to migrate into the MgO matrix during FeCoB deposition [68]. Stress in the interface layer may stabilize the stress-induced UMA in the FeCoB layer at the interface [69]. In general, the FeCoB layer does not possess UMA due to the absence of long-range structural order.

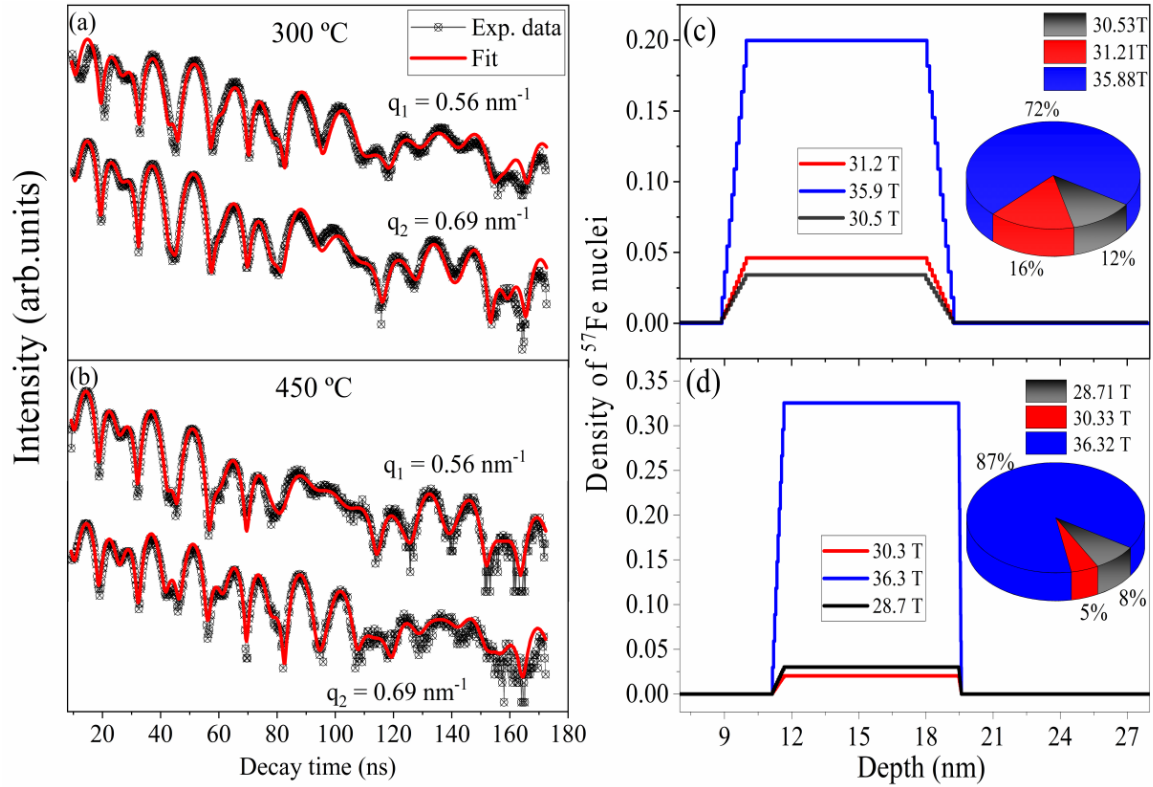

**Figure 7 (a, b).** GINRS time spectra measured at grazing angles ($q_1 = 0.56$ & $q_2 = 0.69$ nm$^{-1}$) before the critical angle of Pt for selected antinode positions of TE$_1$ and TE$_2$ (symbols are the experimental counts and lines are the fit) after annealing at 300 °C and 450 °C. (c, d) The depth distribution of the $^{57}$Fe nuclei with selected types of hyperfine fields.

To make it clearer, contributions of two magnetic components in a single magnetic layer are simulated with a mathematical function [13] by taking Kerr signal K(H), coercive field ($H_c$), squareness ($M_r$) for two magnetic components corresponding to the FeCoB (FeCoB$_{top}$) and interface layer (FeCoB$_{bot}$).

$$K(H) = K(H)_{FeCoB(top)} + K(H)_{FeCoB(bot)}$$

$$K(H) = \frac{2M_s^{top}}{\pi} \arctan\left|\frac{(H \pm H_c^{top})}{H_c^{top}} \tan(\frac{\pi M_r^{top}}{2})\right| + \frac{2M_s^{bot}}{\pi} \arctan\left|\frac{(H \pm H_c^{bot})}{H_c^{bot}} \tan(\frac{\pi M_r^{bot}}{2})\right| \qquad 10.$$

Where $M_r^{top}$ and $M_r^{bot}$ are magnetic remanence and $H_c^{top}$ and $H_c^{bot}$ are coercivity of $FeCoB_{top}$ and $FeCoB_{bot}$ layer, respectively. The values of these parameters are taken by assuming the interface layer is anisotropic (along the easy axis: $M_r^{bot} = 0.96, M_s^{bot} = 0.5$, and $H_c^{bot} = 25$ Oe; along the hard axis: $M_r^{bot} = 0.15$, $M_s^{bot} = 0.5$, and $H_c^{bot} = 2$ Oe) and FeCoB layer isotropic layer ($M_r^{top} = 0.96, M_s^{top} = 0.5$, and $H_c^{top} = 25$ Oe ). The calculated loops along easy and hard directions for both layers are presented in figures 8 (a) and 8(b), respectively.

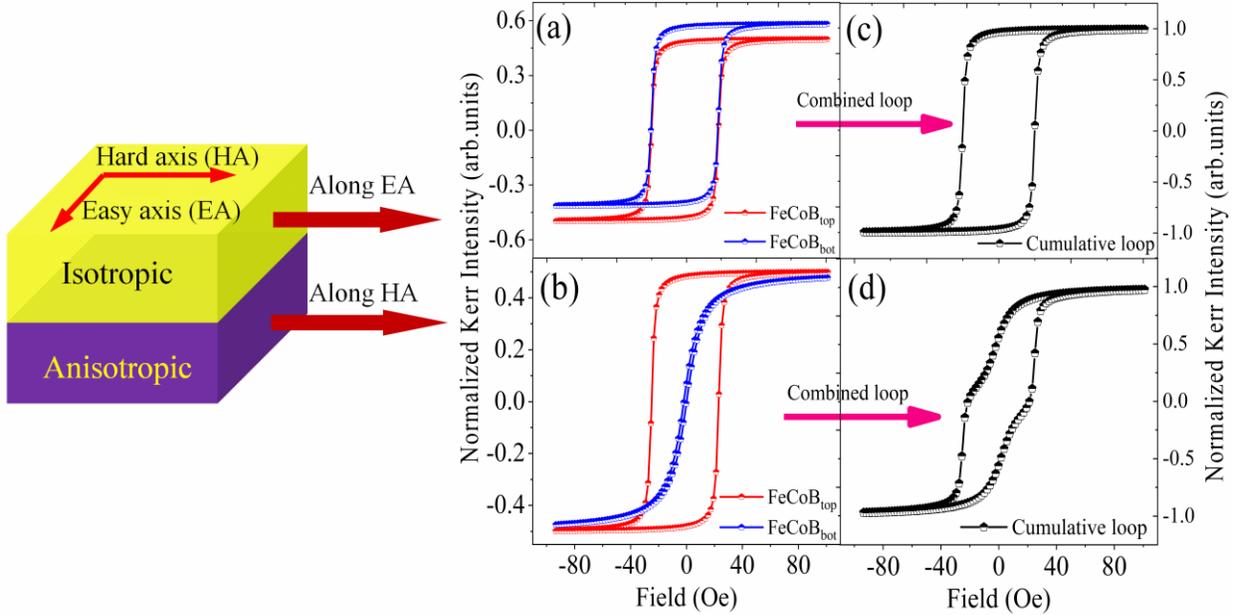

**Figure 8.** Simulated loops along the easy and hard axis of magnetization with (a) and without (b) magnetic anisotropy. In figure (a) blue loop translated upward with respect to the red loop for a clear view. Combined loops along easy (c) and hard (d) directions if both are combined in a single layer.

One of the magnetic components exhibits angular dependence, a square loop along an easy and sheared loop in a hard direction, similar to the Stoner-Wohlfarth model [70] for uniaxial magnetic anisotropy. At the same time, loops are similar in easy and hard directions for the isotropic layer. The combined contribution of both the layer along easy and hard directions is calculated using equation (10) and presented in figure 8 (c) and 8(d). The shape of the combined loop is similar to the loops observed in the reported sample. It suggests that the $FeCoB_{bot}$ layer in the present case exhibits different magnetization processes in the film plane due to UMA and couples differently with the magnetically isotropic $FeCoB_{top}$ layer in different directions. Similar anisotropic coupling between two different magnetically anisotropic components in a single layer is observed and found to result in invariable coupling as a function of angle [13,71,72].

The present understanding revealed that the coupled domain wall motion in $FeCoB_{top}$ and $FeCoB_{bot}$ parts together gives such an unusual hysteresis loop. The rounding-off of the hysteresis loop from easy to hard direction indicates an increasing contribution of the rotation of domain magnetization in the FeCoB$_{bot}$ layer, which is in contrast to the domain wall motion in the FeCoB$_{top}$ layer. Hence, the magnetization process of the combined FeCoB layer is significantly different from either of the loops, resulting in a double-step loop with increased coercivity near the hard axis in the combined hysteresis loop of the FeCoB layer. A similar observation of two-step magnetization reversal in FeCoB/MgO bilayer thin film is reported in reference [13]. The origin of stress-induced magnetic anisotropy in the interfacial FeCoB layer is attributed to the boron diffusion from the bulk FeCoB layer to the MgO layer.

The disappearance of anisotropic loops after annealing at 450 °C is correlated with the nanocrystallization of amorphous FeCoB thin film. Since the density of the crystalline bcc-FeCo phase is higher than that of the amorphous phase, the nano-crystals exert random tensile stress in the amorphous matrix around it and vice versa [73]. Random stress in the crystalline FeCoB film after annealing is expected to overcome the long-range stresses present in the as-deposited film, resulting in the disappearance of magnetic anisotropy. The observed increase in the coercivity and hyperfine field are well correlated with the initiation of crystallization of the amorphous FeCoB phase. Additionally, structural defects and compositional inhomogeneities may be generated after annealing and will also hinder the motion of domain walls and would contribute to the increased coercivity.

## 6. Conclusion

In conclusion, MgO/FeCoB/MgO trilayer structure grown in Pt waveguide has been studied with an aim to study the origin of the unusual angular dependent magnetism. Depth-resolved interface magnetism in the buried structure is resolved by generating XSW in the waveguide. The field-intensity enhancement has been achieved by performing synchrotron radiation-based GINRS at different grazing incident angles at XSW modes TE$_0$, TE$_1$, and TE$_2$ positions. XSW technique is used to elucidate the asymmetry in the magnetism of the two interfaces FeCoB-on-MgO and MgO-on-FeCoB, and the same is correlated with magnetic properties. The combined structural analysis revealed the formation of a high-density layer at the FeCoB-on-MgO interface due to B diffusion from FeCoB into the MgO matrix results in the formation of Fe and Co-rich interface layer. B diffusion into MgO at the bottom interface is attributed to the porousness of the MgO layer, which might have been generated during the deposition of the film. Due to the small size of B compared to the Fe and Co and the high electron affinity of Mg and O, B migrates into the MgO matrix during FeCoB deposition. Stress in the boron-deficient interface layer coupled with the magnetoelastic energy resulted in the preferential orientation of magnetization along the easy axis in the film plane. The observed preferential orientation of moments in the FeCoB$_{bot}$ interface layer and its coupling with the rest of the magnetic layer (FeCoB$_{top}$) cause unusual UMA in this system. The systematic reduction in magnetic anisotropy with temperature is due to the removal of stress and crystallization of FeCoB. The disappearance of anisotropy after annealing at 450 °C is mainly due to the removal of stress and the formation of the crystalline bcc-FeCo phase.

## Acknowledgments

Portion of this research was carried out at the light source PETRA-III of DESY, a member of Helmholtz Association (HGF). Financial support by the Department of Science & Technology (Government of India) (Proposal No. I-20220458) provided within the framework of the India@DESY collaboration is gratefully acknowledged. We would also like to acknowledge Dr. Mukul Gupta and Mr. Yogesh Kumar for the sample preparation.

## References

[1] S. Ikeda, K. Miura, H. Yamamoto, K. Mizunuma, H. D. Gan, M. Endo, S. Kanai, J. Hayakawa, F. Matsukura, and H. Ohno, *A Perpendicular-Anisotropy CoFeB-MgO Magnetic Tunnel Junction*, Nat. Mater. **9**, 721 (2010).

[2] J. Sinha, M. Hayashi, A. J. Kellock, S. Fukami, M. Yamanouchi, H. Sato, S. Ikeda, S. Mitani, S.-H. Yang, S. S. P. Parkin, et al., *Enhanced Interface Perpendicular Magnetic Anisotropy in Ta/CoFeB/MgO Using Nitrogen Doped Ta Underlayers*, Cit. Appl. Phys. Lett. **102**, 113902 (2013).

[3] A. A. Greer, A. X. Gray, S. Kanai, A. M. Kaiser, S. Ueda, Y. Yamashita, C. Bordel, G. Palsson, N. Maejima, S.-H. Yang, et al., *Observation of Boron Diffusion in an Annealed Ta/CoFeB/MgO Magnetic Tunnel Junction with Standing-Wave Hard x-Ray Photoemission*, Appl. Phys. Lett. **101**, 202402 (2012).

[4] B. Dieny and M. Chshiev, *Perpendicular Magnetic Anisotropy at Transition Metal/Oxide Interfaces and Applications*, Rev. Mod. Phys. **89**, (2017).

[5] W. Skowroński, T. Nozaki, Y. Shiota, S. Tamaru, K. Yakushiji, H. Kubota, A. Fukushima, S. Yuasa, and Y. Suzuki, *Perpendicular Magnetic Anisotropy of Ir/CoFeB/MgO Trilayer System Tuned by Electric Fields*, Appl. Phys. Express **8**, (2015).

[6] N. H. Kim, J. Jung, J. Cho, D. S. Han, Y. Yin, J. S. Kim, H. J. M. Swagten, and C. Y. You, *Interfacial Dzyaloshinskii-Moriya Interaction, Surface Anisotropy Energy, and Spin Pumping at Spin Orbit Coupled Ir/Co Interface*, Appl. Phys. Lett. **108**, 142406 (2016).

[7] S. Woo, M. Mann, A. J. Tan, L. Caretta, and G. S. D. Beach, *Enhanced Spin-Orbit Torques in Pt/Co/Ta Heterostructures*, Appl. Phys. Lett. **105**, 212404 (2014).

[8] Y. Ou, C.-F. Pai, S. Shi, D. C. Ralph, and R. A. Buhrman, *Origin of Fieldlike Spin-Orbit Torques in Heavy Metal/Ferromagnet/Oxide Thin Film Heterostructures*, RAPID Commun. Phys. Rev. B

**94**, 140414 (2016).

[9] G. Prenat, K. Jabeur, P. Vanhauwaert, G. Di Pendina, F. Oboril, R. Bishnoi, M. Ebrahimi, N. Lamard, O. Boulle, K. Garello, et al., *Ultra-Fast and High-Reliability SOT-MRAM: From Cache Replacement to Normally-Off Computing*, IEEE Trans. Multi-Scale Comput. Syst. **2**, 49 (2016).

[10] Y. Liu, X. Liu, and J. G. Zhu, *Tailoring the Current-Driven Domain Wall Motion by Varying the Relative Thickness of Two Heavy Metal Underlayers*, IEEE Trans. Magn. **54**, (2018).

[11] M.-H. Nguyen, C.-F. Pai, and K. X. Nguyen, *Cite As*, Appl. Phys. Lett **106**, 222402 (2015).

[12] M. Raju, S. Chaudhary, and D. K. Pandya, *Multi-Jump Magnetic Switching in Ion-Beam Sputtered Amorphous Co 20Fe60B20 Thin Films*, J. Appl. Phys. **114**, (2013).

[13] S. Singh, D. Kumar, M. Gupta, and N. P. Lalla, *Study of Interface Induced Anisotropic Exchange Coupling in Amorphous FeCoB/MgO Bilayer*, J. Alloys Compd. **789**, 330 (2019).

[14] Y. S. Choi, Y. Nagamine, K. Tsunekawa, H. Maehara, D. D. Djayaprawira, S. Yuasa, and K. Ando, *Effect of Ta Getter on the Quality of MgO Tunnel Barrier in the Polycrystalline CoFeB/MgO/CoFeB Magnetic Tunnel Junction*, Appl. Phys. Lett. **90**, 012505 (2007).

[15] D. D. Djayaprawira, K. Tsunekawa, M. Nagai, H. Maehara, S. Yamagata, N. Watanabe, S. Yuasa, Y. Suzuki, and K. Ando, *230% Room-Temperature Magnetoresistance in CoFeB/MgO/CoFeB Magnetic Tunnel Junctions*, Appl. Phys. Lett. **86**, 092502 (2005).

[16] J. C. Read, P. G. Mather, and R. A. Buhrman, *X-Ray Photoemission Study of Co Fe B / Mg O Thin Film Bilayers*, Cit. Appl. Phys. Lett. **90**, 132503 (2007).

[17] J. J. Cha, J. C. Read, R. A. Buhrman, and D. A. Muller, *Spatially Resolved Electron Energy-Loss Spectroscopy of Electron-Beam Grown and Sputtered CoFeBMgOCoFeB Magnetic Tunnel Junctions*, Appl. Phys. Lett. **91**, (2007).

[18] S. S. Mukherjee, F. Bai, D. MacMahon, C. L. Lee, S. K. Gupta, and S. K. Kurinec, *Crystallization and Grain Growth Behavior of CoFeB and MgO Layers in Multilayer Magnetic Tunnel Junctions*, J. Appl. Phys. **106**, 033906 (2009).

[19] S. Pinitsoontorn, A. Cerezo, A. K. Petford-Long, D. Mauri, L. Folks, and M. J. Carey, *Three-Dimensional Atom Probe Investigation of Boron Distribution in Co Fe B / Mg O / Co Fe B Magnetic Tunnel Junctions*, Cit. Appl. Phys. Lett. **93**, 103921 (2008).

[20] Z. Bai, L. Shen, Q. Wu, M. Zeng, J.-S. Wang, G. Han, and Y. P. Feng, *PACS Number(s): 72.25.Mk, 63.20.−e, 66.30.−h, 75*, Phys. Rev. B **87**, 14114 (2013).

[21] Y. Lu, M. Tran, H. Jaffrè, P. Seneor, C. Deranlot, F. Petroff, J.-M. George, B. Lépine, S. Ababou, and G. Jézéquel, *Spin-Polarized Inelastic Tunneling through Insulating Barriers*, (2009).

[22] Y. Jang, C. Nam, K.-S. Lee, B. K. Cho, Y. J. Cho, K.-S. Kim, and K. W. Kim, *Variation in the Properties of the Interface in a CoFeB / MgO / CoFeB Tunnel Junction during Thermal Annealing*,

(2007).

[23] D. A. Stewart, *New Type of Magnetic Tunnel Junction Based on Spin Filtering through a Reduced Symmetry Oxide: FeCo | Mg3B2O6| FeCo*, Nano Lett. **10**, 263 (2010).

[24] W. G. Wang, J. Jordan-Sweet, G. X. Miao, C. Ni, A. K. Rumaiz, L. R. Shah, X. Fan, P. Parsons, R. Stearrett, E. R. Nowak, et al., *In-Situ Characterization of Rapid Crystallization of Amorphous CoFeB Electrodes in CoFeB/MgO/CoFeB Junctions during Thermal Annealing*, Appl. Phys. Lett. **95**, 242501 (2009).

[25] W. G. Wang, C. Ni, A. Rumaiz, Y. Wang, X. Fan, T. Moriyama, R. Cao, Q. Y. Wen, H. W. Zhang, J. Q. Xiao, et al., *Real-Time Evolution of Tunneling Magnetoresistance during Annealing in CoFeB/MgO/CoFeB Magnetic Tunnel Junctions*, Appl. Phys. Lett. **92**, 152501 (2008).

[26] Y. M. Lee, J. Hayakawa, S. Ikeda, F. Matsukura, and H. Ohno, *Giant Tunnel Magnetoresistance and High Annealing Stability in CoFeB/MgO/CoFeB Magnetic Tunnel Junctions with Synthetic Pinned Layer*, Appl. Phys. Lett. **89**, 042506 (2006).

[27] P. V Paluskar, J. J. Attema, G. A. De Wijs, S. Fiddy, E. Snoeck, J. T. Kohlhepp, H. J. M. Swagten, R. A. De Groot, and B. Koopmans, *Spin Tunneling in Junctions with Disordered Ferromagnets*, (2007).

[28] H. Kurt, K. Oguz, T. Niizeki, and J. M. D. Coey, *Giant Tunneling Magnetoresistance with Electron Beam Evaporated MgO Barrier and CoFeB Electrodes*, J. Appl. Phys. **107**, 083920 (2010).

[29] R. P. Cowburn, S. J. Gray, and J. A. C. Bland, *Multijump Magnetic Switching in In-Plane Magnetized Ultrathin Epitaxial Ag/Fe/Ag(001) Films*, Phys. Rev. Lett. **79**, 4018 (1997).

[30] R. P. Cowburn, S. J. Gray, J. Ferré, J. A. C. Bland, and J. Miltat, *Magnetic Switching and In-Plane Uniaxial Anisotropy in Ultrathin Ag/Fe/Ag(100) Epitaxial Films*, J. Appl. Phys. **78**, 7210 (1995).

[31] J. Chen and J. L. Erskine, *(( I Jf*, **68**, 1212 (1992).

[32] Q. F. Zhan, S. Vandezande, K. Temst, and C. Van Haesendonck, *Magnetic Anisotropy and Reversal in Epitaxial Fe/MgO(001) Films*, Phys. Rev. B - Condens. Matter Mater. Phys. **80**, 1 (2009).

[33] Q. F. Zhan, S. Vandezande, C. Van Haesendonck, and K. Temst, *Manipulation of In-Plane Uniaxial Anisotropy in FeMgO (001) Films by Ion Sputtering*, Appl. Phys. Lett. **91**, 22 (2007).

[34] A. T. Hindmarch, C. J. Kinane, M. MacKenzie, J. N. Chapman, M. Henini, D. Taylor, D. A. Arena, J. Dvorak, B. J. Hickey, and C. H. Marrows, *Interface Induced Uniaxial Magnetic Anisotropy in Amorphous CoFeB Films on AlGaAs(001)*, Phys. Rev. Lett. **100**, 1 (2008).

[35] L. Kipgen, H. Fulara, M. Raju, and S. Chaudhary, *In-Plane Magnetic Anisotropy and Coercive Field Dependence upon Thickness of CoFeB*, J. Magn. Magn. Mater. **324**, 3118 (2012).

[36] Z. Hussain, V. R. Reddy, D. Kumar, V. Ganesan, V. Dhamgaye, N. Khantwal, and A. Gupta, *Study of Two Phase Magnetization Reversal in Patterned Cobalt Thin Film*, J. Phys. D. Appl. Phys. **50**,

425001 (2017).

[37] J. C. Read, P. G. Mather, and R. A. Buhrman, *X-Ray Photoemission Study of CoFeB/MgO Thin Film Bilayers*, Appl. Phys. Lett. **90**, 132503 (2007).

[38] G. Tsoy, Z. Janu, M. Novak, F. Soukup, and R. Tichy, *High-Resolution SQUID Magnetometer*, Phys. B Condens. Matter **284–288**, 2122 (2000).

[39] A. W. Pacyna and K. Ruebenbauer, *General Theory of a Vibrating Magnetometer with Extended Coils*, J. Phys. E. **17**, 141 (1984).

[40] M. S. Jamal, P. Gupta, and D. Kumar, *Preferential Alignment of Co Moments at Oxide/Co Interface in Si/Co/Co-Oxide/Cowedge/Pt Structure*, Thin Solid Films **709**, 138246 (2020).

[41] D. Kumar, A. Gupta, P. Patidar, A. Banerjee, K. K. Pandey, T. Sant, and S. M. Sharma, *Interface Induced Perpendicular Magnetic Anisotropy in a Co/CoO/Co Thin-Film Structure: An in Situ MOKE Investigation*, J. Phys. D. Appl. Phys. **47**, (2014).

[42] L. K. Perry, H. Ryan, and R. Gagnon, *Studying Surfaces and Thin Films Using Mössbauer Spectroscopy*, NASSAU 2006 131 (2006).

[43] S. Won, S. B. Saun, S. Lee, S. Lee, K. Kim, and Y. Han, *NMR Spectroscopy for Thin Films by Magnetic Resonance Force Microscopy*, Sci. Rep. **3**, (2013).

[44] A. Gupta, D. Kumar, C. Meneghini, and J. Zegenhagen, *Depth Resolved X-Ray Absorption Fine Structure Study in Magnetic Multilayers Using x-Ray Standing Waves*, J. Appl. Phys. **101**, 09D117 (2007).

[45] M. S. Jamal, P. Gupta, R. Raj, M. Gupta, V. R. Reddy, and D. Kumar, *Structural and Magnetic Asymmetry at the Interfaces of MgO/FeCoB/MgO Trilayer: Precise Study under x-Ray Standing Wave Conditions*, J. Appl. Phys **131**, 235301 (2022).

[46] P. Vishwakarma, M. Nayak, V. R. Reddy, A. Gloskovskii, W. Drube, and A. Gupta, *Standing Wave Hard X-Ray Photoemission Study of the Structure of the Interfaces in Ta/Co2FeAl/MgO Multilayer*, Appl. Surf. Sci. **590**, 153063 (2022).

[47] M. S. Jamal, Y. Kumar, M. Gupta, P. Gupta, I. Sergeev, H. C. Wille, and D. Kumar, *Study of Interface and Its Role in an Unusual Magnetization Reversal in 57FeCoB/MgO Bilayer*, Hyperfine Interact. **242**, 4 (2021).

[48] R. Rüffer and A. I. Chumakov, *Nuclear-Resonance Beamline at ESRF*, Nuovo Cim. Della Soc. Ital. Di Fis. D - Condens. Matter, At. Mol. Chem. Physics, Biophys. **18**, 375 (1996).

[49] A. Gupta, *Depth Resolved Structural Studies in Multilayers Using X-Ray Standing Waves*, Hyperfine Interact. 2005 1601 **160**, 123 (2005).

[50] A. Gupta, M. Gupta, S. Chakravarty, R. Rüffer, H.-C. Wille, and O. Leupold, *Fe Diffusion in Amorphous and Nanocrystalline Alloys Studied Using Nuclear Resonance Reflectivity*, (n.d.).

[51] M. Andreeva, A. Smekhova, R. Baulin, Y. Repchenko, R. Bali, C. Schmitz-Antoniak, H. Wende, I. Sergueev, K. Schlage, and H.-C. Wille, *Evolution of the Magnetic Hyperfine Field Profiles in an Ion-Irradiated Fe60Al40 Film Measured by Nuclear Resonant Reflectivity*, J. Synchrotron Radiation1600-5775 **28**, (2021).

[52] A. G. Khanderao, I. Sergueev, H. C. Wille, and D. Kumar, *Interface Resolved Magnetism at Metal-Organic (Fe/Alq3) Interfaces under x-Ray Standing Wave Condition*, Appl. Phys. Lett. **116**, (2020).

[53] L. G. Parratt, *Surface Studies of Solids by Total Reflection of X-Rays*, Phys. Rev. **95**, 359 (1954).

[54] M. A. Andreeva, *Nuclear Resonant Reflectivity Data Evaluation with the "REFTIM" Program*, Hyperfine Interact **185**, 17 (2008).

[55] B. C. E Johnson, M. S. Ridout, and T. E. Cranshaw, *The Mossbauer Effect in Iron Alloys*, **81**, (1963).

[56] H.-P. Klein, M. Ghafari, M. Ackermann, U. Gonser, and H.-G. Wagner, *CRYSTALLIZATION OF AMORPHOUS METALS*, Nucl. Instruments Methods **199**, 159 (1982).

[57] D. Kumar and A. Gupta, *Evolution of Structural and Magnetic Properties of Sputtered Nanocrystalline Co Thin Films with Thermal Annealing*, J. Magn. Magn. Mater. **308**, 318 (2007).

[58] R. C. O'HANDLEY, *Modern Magnetic Materials: Principles and Applications*, Vol. 13 (1942).

[59] A. S. Dev, A. K. Bera, P. Gupta, V. Srihari, P. Pandit, M. Betker, M. Schwartzkopf, S. V. Roth, and D. Kumar, *Oblique Angle Deposited FeCo Multilayered Nanocolumnar Structure: Magnetic Anisotropy and Its Thermal Stability in Polycrystalline Thin Films*, Appl. Surf. Sci. **590**, (2022).

[60] L. Néel, R. Pauthenet, G. Rimet, and V. S. Giron, *On the Laws of Magnetization of Ferromagnetic Single Crystals and Polycrystals. Application to Uniaxial Compounds*, Cit. J. Appl. Phys. **31**, 27 (1960).

[61] J. Dwivedi, R. Gupta, G. Sharma, M. Gupta, A. Mishra, and A. Gupta, *Annealing Induced Structural Changes in Amorphous Co23Fe60B17 Film on Mo Buffer Layer*, AIP Conf. Proc. **1731**, 1 (2016).

[62] P. Gupta, K. J. Akhila, V. Srihari, P. Svec, S. R. Kane, S. K. Rai, and T. Ganguli, *On the Origin of Magnetic Anisotropy of FeCo(Nb)B Alloy Thin Films: A Thermal Annealing Study*, J. Magn. Magn. Mater. **480**, 64 (2019).

[63] V. Barwal, S. Husain, N. Behera, E. Goyat, and S. Chaudhary, *Growth Dependent Magnetization Reversal in Co 2 MnAl Full Heusler Alloy Thin Films*, Cit. J. Appl. Phys. **123**, 53901 (2018).

[64] M. Mathews, E. P. Houwman, H. Boschker, G. Rijnders, and D. H. A. Blank, *Magnetization Reversal Mechanism in La0.67Sr 0.33MnO3 Thin Films on NdGaO3 Substrates*, J. Appl. Phys. **107**, (2010).

[65] A. T. Hindmarch, A. W. Rushforth, R. P. Campion, C. H. Marrows, and B. L. Gallagher, *Origin of In-Plane Uniaxial Magnetic Anisotropy in CoFeB Amorphous Ferromagnetic Thin Films*, Phys. Rev. B - Condens. Matter Mater. Phys. **83**, 1 (2011).

[66] V. Thiruvengadam, A. Mishra, S. Mohanty, and S. Bedanta, *Anisotropy and Domain Structure in Nanoscale-Thick MoS 2 /CoFeB Heterostructures: Implications for Transition Metal Dichalcogenide- Based Thin Films*, ACS Appl. Nano Mater. (2022).

[67] M. Endo, S. Kanai, S. Ikeda, F. Matsukura, and H. Ohno, *Electric-Field Effects on Thickness Dependent Magnetic Anisotropy of Sputtered MgO/Co40Fe40B20/Ta Structures*, Appl. Phys. Lett. **96**, 212503 (2010).

[68] V. Harnchana, A. T. Hindmarch, M. C. Sarahan, C. H. Marrows, A. P. Brown, and R. M. D. Brydson, *Evidence for Boron Diffusion into Sub-Stoichiometric MgO (001) Barriers in CoFeB/MgO-Based Magnetic Tunnel Junctions*, J. Appl. Phys. **113**, (2013).

[69] D. Kumar, S. Singh, P. Vishawakarma, A. S. Dev, V. R. Reddy, and A. Gupta, *Tailoring of In-Plane Magnetic Anisotropy in Polycrystalline Cobalt Thin Films by External Stress*, J. Magn. Magn. Mater. **418**, 99 (2016).

[70] E. C. Stoner and E. P. Wohlfarth, *A Mechanism of Magnetic Hysteresis in Heterogeneous Alloys*, Royal Society **826**, 599 (1948).

[71] A. T. Hindmarch, C. J. Kinane, C. H. Marrows, B. J. Hickey, M. Henini, D. Taylor, D. A. Arena, and J. Dvorak, *In-Plane Magnetic Anisotropies of Sputtered Co0.7Fe0.3 Films on AlGaAs(001) Spin Light Emitting Diode Heterostructures*, J. Appl. Phys. **101**, 09D106 (2007).

[72] A. T. Hindmarch, D. A. Arena, K. J. Dempsey, M. Henini, and C. H. Marrows, *Influence of Deposition Field on the Magnetic Anisotropy in Epitaxial Co 70 Fe 30 Films on GaAs(001)*, (n.d.).

[73] J. Dwivedi, M. Gupta, V. R. Reddy, A. Mishra, V. Srihari, K. K. Pandey, and A. Gupta, *Effect of Heavy Metal Interface on the Magnetic Behaviour and Thermal Stability of CoFeB Film*, J. Magn. Magn. Mater. **466**, 311 (2018).